\documentclass[twocolumn,preprintnumbers,pre,floatfix,amsmath,amssymb,notitlepage,superscriptaddress]{revtex4-1}
\usepackage{epsfig}
\usepackage{amssymb}
\usepackage{bbm}
\usepackage{indentfirst}
\usepackage{graphicx}
\usepackage{amsmath,amssymb}
\usepackage{bpchem}

\usepackage{color}
\usepackage{tabularx}
\usepackage{multirow}
\usepackage{makecell}
\usepackage{array}
\usepackage{appendix}
\newcommand{\PreserveBackslash}[1]{\let\temp=\\#1\let\\=\temp}
\newcolumntype{C}[1]{>{\PreserveBackslash\centering}p{#1}}
\newcolumntype{R}[1]{>{\PreserveBackslash\raggedleft}p{#1}}
\newcolumntype{L}[1]{>{\PreserveBackslash\raggedright}p{#1}}

\input epsf

\begin{document}

\title{Collective dynamics of heterogeneously and nonlinearly coupled phase oscillators}

\author{Can Xu}
\thanks{xucan@hqu.edu.cn}
\affiliation{Institute of Systems Science and College of Information Science and Engineering, Huaqiao University, Xiamen 361021, China}

\author{Xiaohuan Tang}
\affiliation{School of Physics and Electronic Engineering, Jiangsu Normal University, Xuzhou 221116, China}

\author{Huaping L\"u}
\affiliation{School of Physics and Electronic Engineering, Jiangsu Normal University, Xuzhou 221116, China}

\author{Karin Alfaro-Bittner}
\affiliation{Unmanned Systems Research Institute, Northwestern Polytechnical University, Xi'an 710072, China}
\affiliation{Departamento de F\'isica, Universidad T\'ecnica Federico Santa Mar\'ia, Av. Espa\~na 1680, Casilla 110V, Valpara\'iso, Chile}

\author{Stefano Boccaletti}
\affiliation{Unmanned Systems Research Institute, Northwestern Polytechnical University, Xi'an 710072, China}
\affiliation{Universidad Rey Juan Carlos, Calle Tulip\'an s/n, 28933 M\'ostoles, Madrid, Spain}
\affiliation{Moscow Institute of Physics and Technology, 9 Institutskiy per., Dolgoprudny, 141701 Moscow, Russia}
\affiliation{CNR - Institute of Complex Systems, Via Madonna del Piano 10, I-50019 Sesto Fiorentino, Italy}

\author{Matja{\v z} Perc}
\affiliation{Faculty of Natural Sciences and Mathematics, University of Maribor, Koro{\v s}ka cesta 160, 2000 Maribor, Slovenia}
\affiliation{Department of Medical Research, China Medical University Hospital, China Medical University, Taichung, Taiwan}
\affiliation{Alma Mater Europaea ECM, Slovenska ulica 17, 2000 Maribor, Slovenia}
\affiliation{Complexity Science Hub Vienna, Josefst{\"a}dterstra{\ss}e 39, 1080 Vienna, Austria}

\author{Shuguang Guan}\thanks{sgguan@phy.ecnu.edu.cn}
\affiliation{Department of Physics, East China Normal University, Shanghai 200241, China}

\date{\today}

\newcommand{\WARN}[1]{\textcolor{green}{#1}}
\newcommand{\NOTES}[1]{\textcolor{red}{#1}}

\begin{abstract}
Coupled oscillators have been used to study synchronization in a wide range of social, biological, and physical systems, including pedestrian-induced bridge resonances, coordinated lighting up of firefly swarms, and enhanced output peak intensity in synchronizing laser arrays. Here we advance this subject by studying a variant of the Kuramoto model, where the coupling between the phase oscillators is heterogeneous and nonlinear. In particular, the quenched disorder in the coupling strength and the intrinsic frequencies are correlated, and the coupling itself depends on the amplitude of the mean-field of the system. We show that the interplay of these factors leads to a fascinatingly rich collective dynamics, including explosive synchronization transitions, hybrid transitions with hysteresis absence, abrupt irreversible desynchronization transitions, and tiered phase transitions with or without a vanishing onset. We develop an analytical treatment that enables us to determine the observed equilibrium states of the system, as well as to explore their asymptotic stability at various levels. Our research thus provides theoretical foundations for a number of self-organized phenomena that may be responsible for the emergence of collective rhythms in complex systems.
\end{abstract}

\maketitle

\section{Introduction}
\label{sec:01}

The Kuramoto model is a well-known and widely applicable paradigm for theoretically describing and modeling collective behavior in large ensembles of interacting units~\cite{kuramoto1975in}. Since its conception in 1975, the model has been used to explore synchronization dynamics in numerous social, biological, and physical systems, with examples ranging from arrays of Josephson junctions~\cite{marvel2009invar}, flashing fireflies~\cite{er1991an}, cardiac pacemaker cells~\cite{osaka2017mod}, to economic markets~\cite{may2008comp}. The model has been particularly popular in physics and mathematics, because it often affords analytical insights towards better understanding different types of collective behavior and synchronization transitions that occur in systems consisting of a large number of coupled nonlinear self-sustained oscillators~\cite{strogatz2000from, acebron2005the, pikovsky2015dynamics, Boccaletti2018}.

In its original version, the Kuramoto model elucidates synchronization at the onset of a phase transition that emerges due to the interplay of the intrinsic frequencies of individual oscillators and the global coupling among them. Specifically, the heterogeneity of the natural frequencies of oscillators, which are distributed randomly across the population according to a prescribed probability density, tends to desynchronize the system, while the attracting global coupling between the oscillators opposes this disorder and tends towards synchrony. These two competing forces result in the rich dynamics of the Kuramoto model that has been explored in great detail during the past decades~\cite{xu2020universal}.

In addition to the heterogeneity in natural frequencies, recent work in physics and network science has highlighted the importance of heterogeneous coupling between dynamical units. Namely, the coupling strength between individual oscillators need not be uniform, but rather oscillator-dependent~\cite{hong2011kuramoto, latsenko2013stationary}. Such heterogeneities can be encoded by means of quenched random interactions, which often represent a realistic setup for a wide range of systems and applications that go beyond the traditional identical coupling~\cite{daido1992, daido2000, stiller2000, daido2005}. For example, such heterogeneous interactions have been used to shed light on the collective dynamics of excitatory-inhibitory neurons, or to describe opinion formation amongst conformists and contrarians in sociology~\cite{cb2003, ms2005, 2008}. In terms of theoretical results obtained with heterogeneous coupling between phase oscillators, previous works have reported the emergence of traveling waves, standing waves, $\pi$-states, glass states, cluster states, as well as chimera states~\cite{hong2011con, hong2012mean, kloumann2104phase,peron2021, jafari2021chimera}.

Almost in parallel to coherent phases observed in network dynamics that emerge from heterogeneous coupling, explosive synchronization transitions have also received a lot of attention~\cite{pazo2005, sb2016, souza2019, bick2021}. The latter represent first-order-like phase transitions that switch abruptly between incoherent and coherent states, and they are typically due to correlations in network topology and oscillator dynamics~\cite{gomze2011explosive, xu2015explosive, xu2019syn}. Since abrupt transitions between two stable states underlie many real-life systems, the phenomenon has received ample attention, for example to better understand bistable perception in the brain dynamics~\cite{wang2013brain} and the robustness and optimal functioning of large-scale power grids~\cite{skardal2015control}. More recently, it has also been shown that explosive synchronization transitions can be due to higher-order interactions, where a link connects more than two oscillators, or due to nonlinear coupling, where, for example, the coupling strength between oscillators is proportional to the coherence of the system~\cite{tt2011mul, komarov2015finite, pikovsky2019low, skardal2019abrupt, millan2020exp}. Research has also shown that these two effects can actually be considered equivalent~\cite{xu2020bifurcation2020, xu2021spectrum, xu2021stability}.

But despite the wealth of findings and insights related to the Kuramoto model and its variants, an unexplored question is what novel effects could different coupling patterns have on the macroscopic dynamics of the system. Particularly so if both heterogeneity and nonlinearity are present in the coupling among phase oscillators. In this case, indeed little is known about the possible collective behaviors and the phase transitions between them that may be observed as a result. In this paper, we therefore consider a new variant of the Kuramoto model, which incorporates heterogeneity and nonlinearity in the coupling. More specifically, we consider correlations between the quenched disorder in the coupling strength and the intrinsic frequencies, and we also consider the coupling to be dependent on the amplitude of the mean-filed of the population. As we will show, the interplay between these effects significantly shapes the overall collective dynamics of the system, giving rise to a number of fascinating phenomena of relevance for the emergence of synchronization. Importantly, the studied model is analytically tractable, so that our research provides theoretical foundations, and in fact a natural mechanism, for the spontaneous emergence of various rhythmic states and the bifurcations associated with them in complex systems, and it does so with relatively small changes to system parameters.

The remainder of this paper is organized as follows. In Sec.~\ref{sec:02}, we present the new variant of the Kuramoto model, and we establish the self-consistency approach to determine the long-term macroscopic dynamics. In Sec.~\ref{sec:03}, we carry out a linear stability analysis for the fully locked, two-cluster, states in a special one-dimensional manifold, and we set up a necessary stability criterion for their occurrence. In Sec.~\ref{sec:04}, we extend this stability result to all perturbed directions in the phase space, and we explore the characteristic dynamics of the system due to perturbations. In Sec.~\ref{sec:05}, we investigate the asymptotic stability of the steady states in the infinite limit, and we obtain the associated eigenspectrum by using the Ott-Antonsen reduction. In Sec.~\ref{sec:06}, we investigate the transitions by means of which the system can move between different coherent states. Various types of phase transition towards synchronization are identified, including the explosive synchronization transition, the abrupt irreversible desynchronization transition, and the tiered synchronization transition with or without a vanishing onset. In Sec.~\ref{sec:07}, we summarize our main conclusions and briefly discuss their possible wider implications.

\section{Mathematical model and its stationary solutions}
\label{sec:02}

As noted, we consider an extension of the Kuramoto model, whose governing equations are
\begin{equation}\label{equ:01}
\dot{\theta}_i=\omega_i+\frac{K_if(R)}{N}\sum_{j=1}^N\sin(\theta_j-\theta_i).
\end{equation}
Here $\theta_i$ is the phase of $i-$th oscillator with $i=1, ..., N$. $\{\omega_i\}$ are the natural frequencies selected from a probability density function $g(\omega)$, which is assumed to be symmetric throughout the paper, i.e., the mean frequency is set to zero and $g(\omega)=g(-\omega)$.

Eq.~(\ref{equ:01}) differs from the classical Kuramoto model in that the uniform coupling strength has been replaced by the heterogeneous coupling $K_i$ with a feedback factor $f(R)$, which is a generic function of the amplitude of the Kuramoto order parameter defined by
\begin{equation}\label{equ:02}
Z(t)=R(t)e^{i\Theta(t)}=\frac{1}{N}\sum_{j=1}^{N}e^{i\theta_j}.
\end{equation}
$R(t)$ measures the level of coherence of the system and $\Theta(t)$ denotes the average phase of the population. $K_i$ and $f(R)$ account for the inhomogeneity and nonlinearity of the coupling, respectively. Without loss of generality, we set $f(R)>0$ (which can be made by rescaling the time) and $K_i=K_{N-i}$. As we will show below, the long-term dynamics of this model is significantly richer than the dynamics of the traditional Kuramoto mode with linear homogeneous coupling.

We emphasize that the inherent nonlinearity, quenched random interactions, and the large number of degrees of freedom make it difficult to understand the quantitative dynamics of Eq.~(\ref{equ:01}) with arbitrary choices of $K_i$ and $f(R)$. We therefore consider only the case where the coupling is symmetric around index $i$ of each oscillator. Aside from this constraint, however, we are free to choose $K_i$ and $f(R)$ arbitrarily.

We first focus on a particular case, in which the inhomogeneities of natural frequencies and the couplings are chosen deterministically rather than randomly. Namely, we set $K_i=K\left| \omega_i\right|$, and $f(R)=R^{\beta-1}$ with $K, \beta >0$. In this setting, the randomness is intrinsic to the oscillators themselves rather than to the coupling between them~\cite{wang2011syn, zhang2013exp, bi2016coe,hong2016pha, xu2018origin, xu2016dyn,  xu2019universal, xu2019bifur, yuan2104dyn}. Moreover, $\beta \geq 1$ ($\beta<1$) sets up a positive (negative) feedback between the coupling and the coherence of the system~\cite{rose2007self, gf2007generalized, zhang2015exp, hung2018, zou2020dyn, gao2020reduction, dai2020dis}.

Before proceeding the analysis, we make a few comments of the model.  As mentioned above, the heterogeneous interactions have been used to shed light on the intrinsic dynamics of glass states, or to describe opinion formation amongst conformists and contrarians in sociology.
On the other hand, the natural frequencies, as a typical heterogeneity of the population, are the intrinsic properties of oscillators themselves. Therefore, it is reasonable to establish correlations between the coupling strength and natural frequencies~\cite{latsenko2013stationary, iatsenko2014, hong2016pha}. In all the quoted studies, the coupling strength depends on the natural frequencies implicitly, and the aim is to uncover the traveling waves and glass states in coupled oscillator populations.

In our model, instead, the coupling strength depends on the natural frequencies explicitly, i.e., we set $K_i=K|\omega_i|$ corresponding to a frequency-weighted coupling (FWC). The FWC was proposed by Ref.~\cite{wang2011syn}, and then it was generalized in Ref.~\cite{zhang2013exp, bi2016coe}. In all those studies, it was shown that a FWC is equivalent to the frequency-degree correlations in complex networks reported in Ref.~\cite{gomze2011explosive}. Both models turn out to give rise to Explosive synchronization phenomena~\cite{note3}.

Regarding the nonlinear coupling (NC), where $f(R)$ is a nonlinear function with respect to $R$. Relevant examples include Ref.~\cite{rose2007self, bai2010}, where NC can be achieved by means of phase reduction in a system of Stuart-Landau oscillators. For convenience, one may set $f(R)=R^{\beta-1}$ (a power-law function), which serves as a typical NC. Similar discussions can be found in Ref.~\cite{gf2007generalized, zou2020dyn}. At the same time, recent works in physics and neuroscience highlight the potential importance of higher-order (or group) interactions, e.g., three-way or more, connections that may be organized via higher-order simplexes or a simplicial complex. The effect of such interactions on the dynamics thus represents an important topic of research in the complex system community. More importantly, recent research has also shown that the higher-order interactions  are equivalent to NC~\cite{xu2021stability, note4}.

The definition of the order parameter Eq.~(\ref{equ:02}) allows us to rewrite Eq.~(\ref{equ:01}) as
\begin{equation}\label{equ:03}
\dot{\theta}_i=\omega_i+K|\omega_i|R^{\beta}\sin(\Theta-\theta_i).
\end{equation}
By doing so, Eq.~(\ref{equ:03}) retains the mean-field character, and $K|\omega_i|R^{\beta}$ displays an effective force acting on each phase oscillator $\theta_i$.

Next, we briefly outline the self-consistency approach to analyze the stationary dynamics of  Eq.~(\ref{equ:01}). The key idea for the self-consistent method is to assume that, in the long-term evolution, the macroscopic quantity $R(t)$ is a constant, and $\Theta(t)$ rotates uniformly, i.e., $\Theta(t)=\Omega t+\Theta_0$. Indeed, we may set $\Omega=\Theta_0=0$ by entering into an appropriate rotating frame and shifting initial conditions~\cite{2013pre, note}. Then, Eq.~(\ref{equ:03}) can be reformulated as
\begin{equation}\label{equ:04}
\dot{\theta}_i=\omega_i-q|\omega_i|\sin\theta_i,
\end{equation}
where the auxiliary parameter $q=K R^{\beta}$ is defined to simplify notation.

We can see from Eq.~(\ref{equ:04}) that each individual oscillator exhibits two types of long-term behavior depending on the magnitude of $q$. Specifically, if $q>1$, all oscillators become phase-locked with
\begin{equation}\label{equ:05}
\sin \theta_i=\frac{\omega_i}{q|\omega_i|},
\end{equation}
and
\begin{equation}\label{equ:06}
\cos \theta_i=\sqrt{1-q^{-2}}.
\end{equation}
When $q<1$, all oscillators are drifting, i.e., they rotate nonuniformly on the unit circle and form a stationary distribution given by
\begin{equation}\label{equ:07}
\rho_d(\theta,\omega)=\frac{\rm{sgn}(\omega) \sqrt{\omega^2(1-q^2)}}{2\pi(\omega-q|\omega|\sin \theta)}.
\end{equation}
The sign function $\rm sgn(\omega)$ appearing in Eq.~(\ref{equ:07}) is needed to ensure the positivity of the distribution.

In contrast to the traditional Kuramoto model and its previously studied variants, we find that in our case the coexistence of the phase-locked and the drifting populations can never occur. In other words, the system remains either in the fully phase-locked or the totally drifting state for a fixed value of $q$. The splitting of the system into the fully phase-locked or the completely drifting state allows us to obtain an expression for the order parameter $R$ self-consistently
\begin{equation}\label{equ:08}
Z=R=\left\langle e^{i\theta}\right\rangle_{q<1},
\end{equation}
and
\begin{equation}\label{equ:09}
Z=R=\left\langle e^{i\theta}\right\rangle_{q>1},
\end{equation}
where $\left\langle \cdot \right\rangle$ denotes the average over the population.

Remarkably, the symmetry of the system implies that $\left\langle \sin \theta \right\rangle_{q>1} =\left\langle e^{i\theta} \right\rangle_{q<1} =0$. Therefore, for $q<1$, the only solution to Eq.~(\ref{equ:08}) is $R=q=0$, which corresponds to the incoherent state with the phases scattered uniformly around the unit circle. Conveniently, for $q>1$, the expression for $R$ becomes simply
\begin{equation}\label{equ:10}
R=\cos \theta_i,
\end{equation}
or
\begin{equation}\label{equ:11}
\bigg (\frac{1}{K}\bigg )^{\frac{1}{\beta}}=F(q)=q^{-\frac{1}{\beta}}\sqrt{1-q^{-2}}.
\end{equation}
The derivations presented above reveal three important results. In the first place, the presence of correlations between frequency and coupling results in a universal phase transition towards synchronization, whose equilibrium is independent of the frequency distribution and system size. Secondly, there are no solutions to Eq.~(\ref{equ:11}) for a sufficiently small value of $K$, since $F(q)$ remains finite for $q \in(1,+\infty)$. And thirdly, there exists a critical coupling $K_c$, beyond which the two coherent state solutions exist located at $q<q_c$ and $q>q_c$, respectively. The critical parameter $q_c$ satisfies the condition $F'(q_c)=0$, i.e.,
\begin{equation}\label{equ:12}
q_c=\sqrt{\beta+1},
\end{equation}
which in turn determines the critical coupling $K_c$ and the critical order parameter $R_c$, yielding
\begin{equation}\label{equ:13}
K_c=\frac{(\beta+1)^{\frac{\beta+1}{2}}}{\beta^{\frac{\beta}{2}}},
\end{equation}
and
\begin{equation}\label{equ:14}
R_c=\sqrt{\frac{\beta}{\beta+1}}.
\end{equation}

In what follows, we will prove that the solutions for $R>R_c$ and $q>q_c$ are attractive, and moreover, correspond exactly to the solutions along the $F'(q)<0$ branch.

\section{Stability in the one-dimensional manifold}
\label{sec:03}

We have shown by means of Eqs.~(\ref{equ:05}) and (\ref{equ:06}) that, depending on the sign of the natural frequencies and depending on the intensity of $q$, the oscillator population splits into two populations in the long run at opposite locations. These observations lead us to consider the following question: What is the stability property of the steady states described by Eq.~(\ref{equ:11}) in this manifold? As we will show, this consideration is quite convenient, because the low-dimensional description of the dynamics in Eq.~(\ref{equ:01}) allows us to deduce the associated critical points and gain some intuition about the stability of phase-locked states in the presently studied modified Kuramoto model.

In the fully locked scenario, the system acts like two giant oscillators, labeled as $\theta_{\pm}$, which correspond to natural frequencies $\pm\omega$, respectively. In such a manifold, Eq.~(\ref{equ:01}) reduces to
\begin{equation}\label{equ:15}
\dot{\theta}_+=\omega+\frac{K|\omega|}{2}R^{\beta-1}\sin(\theta_--\theta_+)
\end{equation}
and
\begin{equation}\label{equ:16}
\dot{\theta}_-=-\omega+\frac{K|\omega|}{2}R^{\beta-1}\sin(\theta_+-\theta_-).
\end{equation}
Without loss of generality, we assume that $\omega>0$, thus having $\sin\theta_{\pm}=\pm q^{-1}$ and $\cos\theta_{\pm}=\sqrt{1-q^{-2}}$, and the order parameter becoming
\begin{equation}\label{equ:17}
R=\frac{1}{2}\Big (e^{i\theta_+}+e^{i\theta_-}\Big ).
\end{equation}
Since $\theta_++\theta_-=0$, Eqs.~(\ref{equ:15}) and (\ref{equ:16}) become one-dimensional if we introduce the phase difference $\phi=\theta_+-\theta_-=2\theta_+$, which then yields
\begin{equation}\label{equ:18}
\dot{\phi}=\omega\big [2-K\Phi(\phi)\big ],
\end{equation}
with
\begin{equation}\label{equ:19}
\Phi(\phi)=\cos^{\beta-1}\frac{\phi}{2}\sin\phi.
\end{equation}

In this one-dimensional manifold, the dynamics is relatively simple to analyze. The steady state solution of Eq.~(\ref{equ:18}) requires that
\begin{equation}\label{equ:20}
\frac{2}{K}=\Phi(\phi),\;\phi \in (0,\pi),
\end{equation}
which implies that the coupling strength $K\geq K_c$, such that
\begin{equation}\label{equ:21}
\frac{2}{K}\leq \max \big [\Phi(\phi)\big ].
\end{equation}
We can then show that
\begin{equation}\label{equ:22}
K_c=\frac{2}{\max \big [\Phi(\phi)\big ]}=\frac{2}{\Phi(\phi_0)}=\frac{(\beta+1)^{\frac{\beta+1}{2}}}{\beta^{\frac{\beta}{2}}},
\end{equation}
and that the critical order parameter $R_c$ is given by
\begin{equation}\label{equ:23}
R_c=\cos\frac{\phi_0}{2}=\sqrt{\frac{\beta}{\beta+1}},
\end{equation}
which are exactly the expressions obtained previously in Eqs.~(\ref{equ:13}) and (\ref{equ:14}).

In order to consider the stability of the solutions beyond $K_c$, we impose a small perturbation $\delta \phi$ away from $\phi$ in Eq.~(\ref{equ:20}) and neglect higher order terms of $\delta \phi$. In doing so, we obtain
\begin{equation}\label{equ:24}
\delta \dot{\phi}=-\omega K \Phi'(\phi) \delta \phi,
\end{equation}
from which it follows that the characteristic value of the perturbation in this direction is proportional to $-\Phi'(\phi)$. This means further that the stable (unstable) condition for $\phi$ is equivalent to $\Phi'(\phi)>0$ [$\Phi'(\phi)<0$]. To corroborate this conclusion, we note that
\begin{equation}\label{equ:25}
\Phi'(\phi)=\frac{1}{2}\cos^{\beta-1}\frac{\phi}{2}\Big (\cos^2 \frac{\phi}{2}-\beta \sin^2 \frac{\phi}{2} \Big ),
\end{equation}
which can be simplified significantly by taking into account Eqs.~(\ref{equ:17}) and (\ref{equ:20}), thus obtaining
\begin{equation}\label{equ:26}
\Phi'(\phi)=\frac{1}{2}R^{\beta-1}\big [R^2(\beta +1)-\beta\big ].
\end{equation}
From this we find that $R>R_c$ leads to $\Phi'(\phi)>0$, thereby determining a stable branch of the solution, while $R<R_c$ leads to $\Phi'(\phi)<0$, corresponding to an unstable solution.

Based on the above treatment, we conclude that the steady state solution along the branch $q>q_c$ (or $F'(q)<0$) is stable at least in this special perturbed direction. However, the stationary solution along the branch $q<q_c$ (or $F'(q)>0$) is unstable in all perturbed directions, because it has a positive characteristic value of the perturbation even in the synchronized manifold.

In the subsequent two sections, we will show that these stability criteria can be extended to any perturbation direction, and this regardless of whether $N$ is finite or infinite.

\section{Stability in the $N$-dimensional manifold}
\label{sec:04}

It is worth pointing out that what is unusual about the currently studied system is that the dynamics of Eq.~(\ref{equ:04}) naturally entrains the oscillators into two groups with equal numbers and with opposite positions. To better understand the nature of these fixed points, we restrict our attention to their asymptotic stability in the $N$-dimensional manifold. We will show that the solutions obtained with $R>R_c$ or $q>q_c$ are indeed stable in all the perturbed directions, while the solutions obtained with $R<R_c$ or $q<q_c$ are not.

For the locked states $(\theta_1, \theta_2... \theta_N)$, let $\theta_i(t)=\theta_i+x_i(t)$ and $\mathbf{x}=(x_1, x_2... x_N)$ be the small perturbations. The linearization of the dynamics described by Eq.~(\ref{equ:01}) is given by the following equation~\cite{mirollo2005the}
\begin{equation}\label{equ:27}
\dot{\mathbf{x}}=\mathbf{J}\mathbf{x},
\end{equation}
where $\mathbf{J}$ is the Jacobian matrix, whose entries are
\begin{equation}\label{equ:28}
\begin{split}
J_{ij}=&\frac{\partial\dot{\theta}_i}{\partial \theta_j}=\frac{\partial}{\partial \theta_j}\bigg [\omega_i+\frac{K|\omega_i|}{N}R^{\beta-1}\sum_{k=1}^{N}\sin(\theta_k-\theta_i)\bigg ]\\
=&\frac{\partial R^{\beta-1}}{\partial \theta_j} \frac{K|\omega_i|}{N} \sum_{k=1}^{N}\sin(\theta_k-\theta_i)\\
&+
 \frac{K|\omega_i|}{N} R^{\beta-1}\sum_{k=1}^{N}\frac{\partial \sin(\theta_k-\theta_i)}{\partial \theta_j}\\
=&|\omega_i|\bigg [\frac{\beta q}{NR}\sin \theta_i\sin\theta_j+\frac{q}{NR}\cos \theta_i\cos\theta_j-q\cos\theta_i\delta_{ij}\bigg ].
\end{split}
\end{equation}
We note that the rotational invariance of the dynamics described by Eq.~(\ref{equ:01}) implies that $\mathbf{J}$ always has a trivial eigenvalue $\lambda=0$ corresponding to an eigenvector $(1, 1... 1)$. To verify that this indeed holds, we point out that the rows of $\mathbf{J}$ all sum to $0$, i.e.,
\begin{equation}\label{equ:29}
\sum_{j=1}^{N}J_{ij}=|\omega_i|q \cos\theta_i\bigg (\frac{\sum_{j=1}^{N}\cos \theta_j}{NR}-\sum_{j=1}^{N}\delta_{ij}\bigg )=0,
\end{equation}
where we have taken into account $\sum_{j=1}^{N}\sin \theta_j=0$ and $\sum_{j=1}^{N}\cos \theta_j=NR$. Aside from $\lambda=0$, the stable condition for the locked states requires that the remaining $N-1$ eigenvalues are all negative.

Before a more in-depth exploration of the eigenvalues of $\mathbf{J}$, we here provide an intuitive argument. We first express
\begin{equation}\label{equ:30}
J_{ij}=|\omega_i|\bigg (\frac{\beta q}{NR}s_is_j+\frac{q}{NR}c_ic
_j-qc_i\delta_{ij}\bigg ),
\end{equation}
where $\mathbf{c}=(c_1, c_2... c_N)$ with $c_i=\cos\theta_i$ and $\mathbf{s}=(s_1, s_2... s_N)$ with $s_i=\sin\theta_i$. Hence, the stability condition is equivalent to $\mathbf{J x  x}<0$ for all $\mathbf{x} \in \mathbb{R}^N$ with $\|\mathbf x \| \neq 0$, and $\mathbf{x}$ is orthogonal to $(1, 1... 1)$. That is, $\sum_{i=1}^{N}x_i=0$, and we thus have

\begin{equation}\label{equ:31}
\mathbf{J x x}=\frac{\beta q}{NR} \sum_{i=1}^{N}|\omega_i|s_ix_i\sum_{j=1}^{N}s_jx_j-q\sum_{i=1}^{N}|\omega_i|c_ix_i^2.
\end{equation}
Following Ref.~\cite{mirollo2005the}, it becomes apparent that $\mathbf{J}$ has at most one positive eigenvalue. If there are two or more positive eigenvalues for $\mathbf{J}$, the span of the corresponding eigenvectors defines at least a two-dimensional subspace, on which $\mathbf{J x x}>0$. On the other hand, one can find a nonzero vector $\mathbf{x}$ on a space that is orthogonal to $\mathbf{s}$, but from the identity above, we have $\mathbf{J x x}<0$, which thus leads to a contradiction.

To explore the eigenvalues of $\mathbf{J}$ more conveniently and in further detail, we introduce a change of coordinates in order to partially diagonalize the system. In particular, if $\mathbf{y}=(y_1, y_2... y_N)$ with $y_i=\sqrt{|\omega_i|c_i}x_i$, we obtain
\begin{equation}\label{equ:32}
\mathbf{J x x}=\frac{\beta q}{NR}\mathbf{(u\cdot y)(v \cdot y)}- q\|\mathbf{y}\|^2.
\end{equation}
The vectors $\mathbf{u}=(u_1, u_2... u_N)$ with $u_i=\sqrt{\frac{|\omega_i|}{c_i}}s_i$, and $\mathbf{v}=(v_1, v_2... v_N)$ with $v_i=\frac{s_i}{\sqrt{|\omega_i|c_i}}$, are introduced to ease notation, and their product is
\begin{equation}\label{equ:33}
\mathbf{u \cdot v}=\sum_{i=1}^{N}\frac{s_i^2}{c_i}=N\frac{q^{-2}}{\sqrt{1-q^{-2}}}.
\end{equation}

In this new coordinate system, we can express
\begin{equation}\label{equ:34}
\mathbf{y}=a\mathbf{ \frac{u}{\| u \|}}+b\mathbf{\frac{v}{\|v\|}}+\mathbf{y}^{\perp},
\end{equation}
with $a$ and $b$ being the projections to the unit vector $\mathbf{ \frac{u}{\| u \|}}$ and $\mathbf{\frac{v}{\|v\|}}$, respectively. $\mathbf{y}^{\perp}$ is the remaining vector, such that $\mathbf{y^{\perp}\cdot u=y^{\perp}\cdot v}=0$. We thus have
\begin{equation}\label{equ:35}
\begin{split}
\mathbf{J x x}=&\big (a^2+b^2\big )\bigg (\frac{\beta q^{-1}}{R^2}-q\bigg )-q\|\mathbf{y}^{\perp}\|^2\\
&+ab\bigg [\frac{\beta q}{NR}\|\mathbf{u}\| \|\mathbf {v}\|+\bigg (\frac{\beta q^{-3}}{R^3}-\frac{2q^{-1}}{R}\bigg )\frac{N}{\|\mathbf{u}\| \|\mathbf {v}\|}\bigg ].
\end{split}
\end{equation}
In particular, for $b=0$, the term around $ab$ vanishes, and the identity above becomes a quadratic form. Since the coefficient $a$ could be arbitrarily small, the stability condition $\mathbf{J x x}<0$ requires that
\begin{equation}\label{equ:36}
\beta<R^2q^2,
\end{equation}
which again recovers the necessary stability condition obtained in Eqs.~(\ref{equ:22}) and (\ref{equ:23}).

To proceed with extracting more details about the eigenvalues of $\mathbf{J}$, we now compute its characteristic polynomial. We therefore express $\mathbf{J}$ in the matrix form
\begin{equation}\label{equ:37}
\mathbf{J=W(C-D)},
\end{equation}
where $\mathbf W$ and $\mathbf D$ are the diagonal matrices with the entries $|\omega_i|$ and $qc_i$ along the diagonal, respectively. The matrix $\mathbf C$ is given by
\begin{equation}\label{equ:38}
C_{ij}=\frac{\beta q}{NR}s_is_j+\frac{q}{NR}c_ic_j.
\end{equation}
The characteristic polynomial is
\begin{equation}\label{equ:39}
\begin{split}
P(\lambda)&=\rm det (\mathbf{J}-\lambda \mathbf{I})\\
&=\rm det\big [\mathbf{W(C-D)}-\lambda \mathbf I\big ]\\
&=\rm det\big [\mathbf{W}(\mathbf{C-D-W}^{-1}\lambda)\big ]\\
&=\rm det \Big \{ \mathbf W(\mathbf{D+W^{-1}}\lambda)\big [(\mathbf D+\lambda \mathbf W^{-1})^{-1}\mathbf C-\mathbf I\big ]\Big \}\\
&=\prod_{i=1}^{N}|\omega_i|\Big (qc_i+\frac{\lambda}{|\omega_i|}\Big )\rm det \big [(\mathbf D+\lambda \mathbf W^{-1})^{-1}\mathbf C-\mathbf I \big ].
\end{split}
\end{equation}

The key task from here on is to determine the second term above. We recall that the rank of $\mathbf{C}$ is only two, and $\rm det\big [(\mathbf D+\lambda \mathbf W^{-1})^{-1}\big ]\neq 0$. The same is true for the matrix $(\mathbf D+\lambda \mathbf W^{-1})^{-1}\mathbf C$. Inspired by this property, we choose an orthogonal basis $\big (\mathbf{\frac{c}{\|c\|},\frac{s}{\|s\|}}\big )$ and  extend it into $\mathbb{R}^N$. The remaining $N-2$ vectors span the kernel of $(\mathbf D+\lambda \mathbf W^{-1})^{-1}\mathbf C$. Therefore, the matrix $(\mathbf D+\lambda \mathbf W^{-1})^{-1}\mathbf C$ if restricted to the orthogonal basis $(\mathbf{\frac{c}{\|c\|},\frac{s}{\|s\|}})$ reduces to a $2\times2$ matrix given by
\begin{equation}\label{equ:40}
(\mathbf D+\lambda \mathbf W^{-1})^{-1}\mathbf C=\left(
\begin{array}{cccc}
Q_{11}(\lambda) &Q_{12}(\lambda)\\[8pt]
Q_{21}(\lambda)&Q_{22}(\lambda)
\end{array}\right),
\end{equation}
where its entries are defined as
\begin{equation}\label{equ:41}
\begin{split}
Q_{11}(\lambda)&=\mathbf{\frac{c}{\|c\|}\cdot( D+\lambda W^{-1})^{-1} C\cdot\frac{c}{\|c\|}}\\
&=\frac{q}{NR}\sum_{i=1}^N\frac{|\omega_i|c_i^2}{\lambda+|\omega_i|qc_i},
\end{split}
\end{equation}
\begin{equation}\label{equ:42}
\begin{split}
Q_{12}(\lambda)&=\mathbf{\frac{c}{\|c\|}\cdot( D+\lambda W^{-1})^{-1} C\cdot\frac{s}{\|s\|}}\\
&=\frac{\beta q^{-1}}{R\mathbf{\|c\| \|s\|}}\sum_{i=1}^N\frac{|\omega_i|s_ic_i}{\lambda+|\omega_i|qc_i},
\end{split}
\end{equation}
\begin{equation}\label{equ:43}
\begin{split}
Q_{21}(\lambda)&=\mathbf{\frac{s}{\|s\|}\cdot( D+\lambda W^{-1})^{-1} C\cdot\frac{c}{\|c\|}}\\
&=\frac{qR}{\mathbf{\|c\| \|s\|}}\sum_{i=1}^N\frac{|\omega_i|s_ic_i}{\lambda+|\omega_i|qc_i},
\end{split}
\end{equation}
\begin{equation}\label{equ:44}
\begin{split}
Q_{22}(\lambda)&=\mathbf{\frac{s}{\|s\|}\cdot( D+\lambda W^{-1})^{-1} C\cdot\frac{s}{\|s\|}}\\
&=\frac{\beta q}{NR}\sum_{i=1}^N\frac{|\omega_i|s_i^2}{\lambda+|\omega_i|qc_i}.
\end{split}
\end{equation}

We note further that the terms $Q_{12}(\lambda)$ and $Q_{21}(\lambda)$ are in fact all zero, because the natural frequencies are grouped into plus-minus pairs, i.e., $\omega_i=-\omega_{N-i}$. This symmetrical case significantly simplifies the characteristic polynomial into
\begin{equation}\label{equ:45}
P(\lambda)=\prod_{i=1}^{N}|\omega_i|\Big (qc_i+\frac{\lambda}{|\omega_i|}\Big )\big [1-Q_{11}(\lambda)\big ]\big [1-Q_{22}(\lambda)\big ].
\end{equation}
The eigenvalues $\lambda$ corresponding to $P(\lambda)=0$ are therefore determined by
\begin{equation}\label{equ:46}
\bigg (\frac{1}{K}\bigg )^{\frac{1}{\beta}}=H_c(\lambda)=\frac{q^{\frac{\beta-1}{\beta}}}{N}\sum_{i=1}^N\frac{|\omega_i|(1-q^{-2})}{\lambda+|\omega_i|\sqrt{q^2-1}},
\end{equation}
and
\begin{equation}\label{equ:47}
\bigg (\frac{1}{K}\bigg )^{\frac{1}{\beta}}=H_s(\lambda)=\frac{\beta q^{\frac{\beta-1}{\beta}}}{N}\sum_{i=1}^N\frac{|\omega_i|q^{-2}}{\lambda+|\omega_i|\sqrt{q^2-1}}.
\end{equation}
For convenience, we order the natural frequencies so that $|\omega_1|\leq|\omega_2|\leq...\leq |\omega_N|$.

The obtained eigenvalues merit three important comments. In the first place, we note that $\lambda=0$ is a trivial solution of Eq.~(\ref{equ:46}) because $K^{-\frac{1}{\beta}}=H_c(0)$  is precisely the self-consistent equation described by Eq.~(\ref{equ:11}). As already mentioned, this property stems from the rotational symmetry of the dynamical equation. Secondly, if $\lambda<-|\omega_N|\sqrt{q^2-1}$, both functions $H_c(\lambda)$ and $H_s(\lambda)$ are negative. This implies that the characteristic Eqs.~(\ref{equ:46})-(\ref{equ:47}) have no roots in the region $\lambda\in(-\infty,-|\omega_N|\sqrt{q^2-1})$. And lastly, $\lim_{\lambda \xrightarrow{0^{\mp}} -|\omega_i|\sqrt{q^2-1}}H_{c(s)}(\lambda)=\pm\infty$, which reflects that, away from their poles, the functions $H_c(\lambda)$ and $H_s(\lambda)$ are strictly decreasing. Accordingly, the characteristic equations must each have exactly one root between any two consecutive poles $(-|\omega_{i+1}|\sqrt{q^2-1},-|\omega_i|\sqrt{q^2-1})$.

Since functions $H_c(\lambda)$ and $H_s(\lambda)$ are decreasing for $\lambda \in (-|\omega_1|\sqrt{q^2-1},+\infty)$, and $\lim_{\lambda \to +\infty}H_{c(s)}(\lambda)=0$, the remaining one eigenvalue is completely determined by the difference $\Delta$ defined by
\begin{equation}\label{equ:48}
\Delta=H_s(0)-\bigg (\frac{1}{K}\bigg )^{\frac{1}{\beta}}.
\end{equation}
Specifically, $\Delta<0$ implies that there is a negative root to Eq.~(\ref{equ:47}) in the region $\lambda \in (-|\omega_1|\sqrt{q^2-1},0)$. On the other hand, for $\Delta>0$, a positive root to Eq.~(\ref{equ:47}) exists in the region $\lambda \in (0, +\infty)$. By combining this with the self-consistency equation, we obtain the elegant identity
\begin{equation}\label{equ:49}
\Delta=\beta q F'(q),
\end{equation}
which directly relates the stability criterion for the locked states to the shape of the self-consistent equation $F(q)$. Namely, if $F'(q)>0$ the steady state solutions are unstable, whereas if $F'(q)<0$ the steady state solutions are attractive and stable, which completes the proof.

\section{Stability in the infinite limit}
\label{sec:05}

Next, to better understand the dynamics that emerges from the above considerations, we study the system in the thermodynamical limit. In particular, we investigate two aspects of the asymptotic stability of the equilibrium states, for which we reduce the system to a low-dimensional equation that governs the long-term macroscopic dynamics. Under the limit $N \to \infty$, the dynamical system is equivalent to the continuity equation for the probability density function $\rho(\theta, \omega, t)$
\begin{equation}\label{equ:50}
\frac{\partial\rho}{\partial t}+\frac{\partial(\rho v)}{\partial \theta}=0.
\end{equation}
Here the velocity $v(\theta, \omega, t)$ is given by
\begin{equation}\label{equ:51}
v(\theta, \omega, t)=\omega+\frac{K|\omega|}{2i}R^{\beta-1}\big (Ze^{-i\theta}-\bar{Z}e^{i\theta}\big ),
\end{equation}
and the complex order parameter $Z(t)$ becomes
\begin{equation}\label{equ:52}
Z(t)=\int_{-\infty}^{+\infty} \int_{0}^{2\pi}e^{i\theta} \rho(\theta,\omega,t)g(\omega) d\omega d\theta.
\end{equation}

By applying the Ott-Antonsen ansatz~\cite{otto2008low, otto2009long}, the probability density function can be expressed as
\begin{equation}\label{equ:53}
\rho(\theta, \omega, t)=\frac{g(\omega)}{2\pi} \bigg [1+\sum_{n=1}^{\infty}\bar{\alpha}^ne^{in\theta}+\sum_{n=1}^{\infty}\alpha^ne^{-in\theta} \bigg ],
\end{equation}
where the variable $\alpha(\omega,t)$ is a complex valued function with $|\alpha(\omega,t)|\leq 1$, and the bar denotes the complex conjugate. This consideration underlines the Poisson kernel that is dynamically invariant, as long as $\alpha(\omega,t)$ evolves according to
\begin{equation}\label{equ:54}
\frac{d\alpha}{dt}=i\omega \alpha+\frac{K|\omega|R^{\beta-1}}{2}\big (Z-\bar{Z}\alpha^2 \big ).
\end{equation}
Meanwhile, $Z(t)$ reduces to
\begin{equation}\label{equ:55}
Z(t)=\int_{-\infty}^{+\infty} g(\omega) \alpha(\omega,t)d\omega =\hat{g}\alpha(\omega,t),
\end{equation}
where we have defined the operator $\hat{g}$ to denote the integral with respect to $\omega$. In this setting, the quantity $\alpha(\omega, t)$ can be interpreted as the local order parameter that is formed by the oscillators with frequencies near $\omega$.

With respect to Eq.~(\ref{equ:54}), we are interested in the steady states $\dot{\alpha}=0$, which yields two-composed solutions
\begin{equation}\label{equ:56}
\alpha_0(\omega)=\left\{
\begin{array}{rl}
i\rm sgn(\omega)\frac{1-\sqrt{1-q^2}}{q},& q<1,\\[8pt]
\sqrt{1-q^{-2}}+i\frac{\omega}{q|\omega|},& q>1.
\end{array} \right.
\end{equation}
We note that the first branch of Eq.~(\ref{equ:56}) corresponds to the drifting population, whereas the second branch corresponds to the phase-locked case. Other possible steady state solutions of $\alpha_0(\omega)$ are ruled out by the constraints $|\alpha_0(\omega)| \leq 1$ and $ R\geq 0$~\cite{omel2013bifurcation}.

To perform the linear stability analysis of $\alpha_0(\omega)$, we introduce a small perturbation to it. Namely, let
\begin{equation}\label{equ:57}
\alpha(\omega, t)=\alpha_0(\omega)+\epsilon\eta(\omega,t),
\end{equation}
with $\epsilon$ being the perturbation magnitude and $\eta(\omega, t)$ the perturbation function. The order parameter $Z(t)$ under perturbation becomes
\begin{equation}\label{equ:58}
Z(t)=R+\epsilon\hat{g}\eta(\omega,t).
\end{equation}
Taking into account $R^{\beta-1}=Z^{\frac{\beta-1}{2}}\cdot\bar{Z}^{\frac{\beta-1}{2}}$, substituting Eqs.~(\ref{equ:57}) and (\ref{equ:58}) into  Eq.~(\ref{equ:54}), and neglecting the higher order terms of $\epsilon$, we get the linear evolution equation for the perturbation
\begin{equation}\label{equ:59}
\frac{d \eta(\omega, t)}{d t}=c(\omega) \eta(\omega, t)+a(\omega)\hat{g} \eta(\omega,t)+b(\omega)\hat{g} \bar{\eta}(\omega,t),
\end{equation}
where
\begin{equation}\label{equ:60}
c(\omega)=\left\{
\begin{array}{rl}
i\omega\sqrt{1-q^2},&q<1,\\[8pt]
-|\omega|\sqrt{q^2-1},&q>1,
\end{array} \right.
\end{equation}
and
\begin{equation}\label{equ:61}
a(\omega)=\frac{|\omega|q}{2R}\bigg (\frac{\beta+1}{2}-\alpha_0^2\frac{\beta-1}{2}\bigg ),
\end{equation}
\begin{equation}\label{equ:612}
b(\omega)=\frac{|\omega|q}{2R}\bigg (\frac{\beta-1}{2}-\alpha_0^2\frac{\beta+1}{2}\bigg ).
\end{equation}

Since $\eta(\omega,t)$ is a complex valued function, we need to switch to a system of two equations containing $\bar{\eta}(\omega,t)$. Thus, if $\mathbf{V}=(\eta,\bar{\eta})$, the $\mathbf{V}$ evolves according to
\begin{equation}\label{equ:63}
\frac{d\mathbf V}{d t}=\mathbf{MV+P}\hat{g}\mathbf V,
\end{equation}
where the matrix $\mathbf M$ is a multiplication operator
\begin{equation}\label{equ:64}
\mathbf{M}=\left(
\begin{array}{cccc}
c(\omega)&0\\[8pt]
0&\bar{c}(\omega)
\end{array}\right),
\end{equation}
and $\mathbf P$ is the coefficient matrix given by
\begin{equation}\label{equ:65}
\mathbf{P}=\left(
\begin{array}{cccc}
a(\omega)&b(\omega)\\[8pt]
\bar{b}(\omega)&\bar{a}(\omega)
\end{array}\right).
\end{equation}

\begin{figure*}
\centering
\footnotesize
\includegraphics[width=\linewidth]{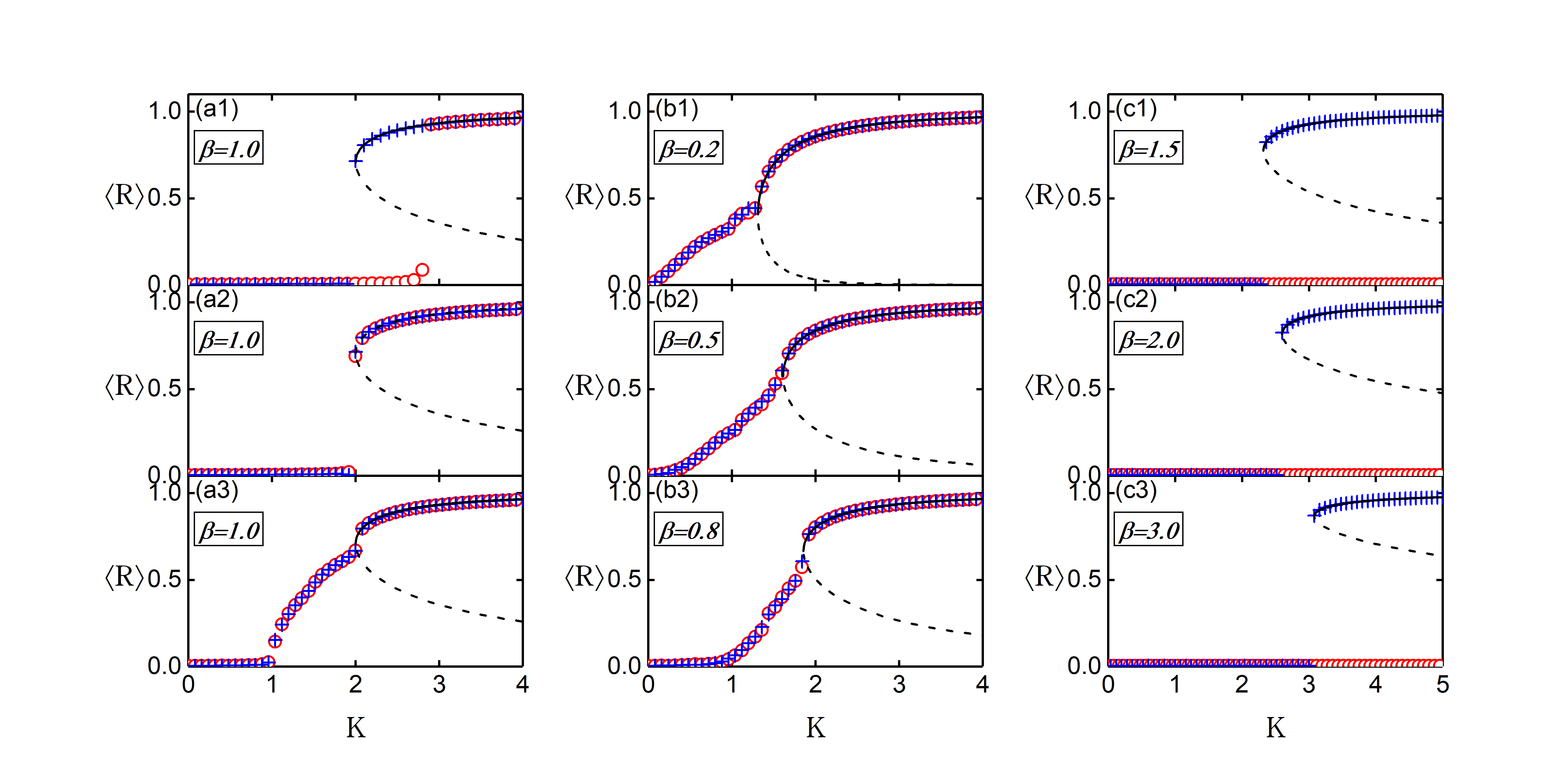}
\caption{Phase diagrams of the time-averaged order parameter $\left\langle R\right\rangle$ versus the global coupling strength $K$, as obtained for different values of the exponent $\beta$. The panels show results for $\beta=1$ (left column), $\beta <1$ (middle column), and $\beta>1$ (right column). For (a1)-(a3) we have used $\beta=1$ and $g(\omega)=\frac{\Delta}{2\pi}\Big [\frac{1}{(\omega-\omega_0)^2+\Delta^2}+\frac{1}{(\omega+\omega_0)^2+\Delta^2}\Big ]$, $K_f=4/\sqrt{1+(\omega_0/ \Delta)^2}$, $K_c=2$, and $R_c=\sqrt{2}/2$. In (a1) $\omega_0=\Delta=0.1$ and $K_f=2\sqrt{2}>K_c$, which leads to a first-order-like phase transition. In (a2) $\omega_0=\sqrt{3}/10$, $\Delta=0.1$, and $K_f=2=K_c$, which, on the other hand, leads to a hybrid phase transition. In (a3) $\omega_0=\sqrt{15}/10$, $\Delta=0.1$, and $K_f=1<K_c$, for which the system undergoes a tiered phase transition, where the oscillatory state arises in the region $K \in (1,2)$. For (b1)-(b3) we have used $\beta<1$, $g(\omega)=1/2$ with $\omega \in(-1,1)$, $K_f=0$, and $K_c=(\beta+1)^{\frac{\beta+1}{2}}\beta^{-{\frac{\beta}{2}}}$. Under these conditions the system undergoes a tiered phase transition with a vanishing onset from the oscillatory state to the two-cluster state. Parameter values were as follows: (b1) $\beta=0.2$, $K_c=1.31$, $R_c=0.41$; (b2) $\beta=0.5$, $K_c=1.61$, $R_c=0.58$; (b3) $\beta=0.8$, $K_c=1.86$, $R_c=0.67$. For (c1)-(c3) we have used $\beta>1$,$g(\omega)=1/\sqrt{2\pi}e^{-\omega^2/2}$, $K_f=+\infty$, and $K_c=(\beta+1)^{\frac{\beta+1}{2}}\beta^{-{\frac{\beta}{2}}}$, leading to an irreversible abrupt desynchronization transition. Parameter values were as follows: (c1) $\beta=1.5$, $K_c=2.32$, $R_c=0.77$; (c2) $\beta=2$, $K_c=2.60$, $R_c=0.82$; (c3) $\beta=3$, $K_c=3.08$, $R_c=0.87$. For the simulations, we have used $N=100000$, $t=950$ with time step $\Delta t=0.01$, and $T=50$. The red circles and the blue crosses represent the forward and backward continuations, respectively. The black lines depict theoretical predictions obtained with the mean-field approach, whereby solid and dash lines depict stable and unstable branches, respectively.}
\label{fig:01}
\end{figure*}

Finally, to conclude the linear stability analysis, let $\lambda \in \mathbb C$, and
\begin{equation}\label{equ:66}
\frac{d\mathbf V}{d t}=\mathbf{MV+P}\hat{g}\mathbf V=\lambda \mathbf V.
\end{equation}
The steady state solution $\mathbf V$ is then solved as
\begin{equation}\label{equ:67}
\mathbf{V=\big (\lambda I-M\big )}^{-1}\mathbf P\hat{g}\mathbf V.
\end{equation}
By applying the integral operator $\hat{g}$ to both sides of Eq.~(\ref{equ:67}), we have the linear equations for the vector $\hat{g}\mathbf V$
\begin{equation}\label{equ:68}
\big (\mathbf{T-I}\big )\hat{g}\mathbf V=0,
\end{equation}
where the matrix $\mathbf T$ is defined by
\begin{equation}\label{equ:69}
\begin{split}
\mathbf T&=\hat{g}\big (\lambda \mathbf I-\mathbf M\big )^{-1}\mathbf P\\
&=\left(
\begin{array}{cccc}
\hat{g}\frac{a(\omega)}{\lambda-c(\omega)}&\hat{g}\frac{b(\omega)}{\lambda-c(\omega)}\\[8pt]
\hat{g}\frac{\bar{b}(\omega)}{\lambda-\bar{c}(\omega)}&\hat{g}\frac{\bar a (\omega)}{\lambda-\bar{c}(\omega)}
\end{array}\right)\\[8pt]
&=\left(
\begin{array}{cccc}
T_{11}(\lambda) &T_{12}(\lambda)\\[8pt]
T_{21}(\lambda)&T_{22}(\lambda)
\end{array}\right).
\end{split}
\end{equation}
Therefore, the nontrivial solution of $\hat{g}\mathbf V$ in Eq.~(\ref{equ:68}) requires that
\begin{equation}\label{equ:70}
\rm det\big (\mathbf T-\mathbf I\big )=0.
\end{equation}

For real $\lambda$, we have $T_{11}(\lambda)=\bar{T}_{22}(\lambda)$, $T_{12}(\lambda)=\bar{T}_{21}(\lambda)$, and it is possible to show that, for a symmetric $g(\omega)$, $T_{11}$ and $T_{12}$ are real. Hence, $T_{11}(\lambda)=T_{22}(\lambda)$ and $T_{12}(\lambda)=T_{21}(\lambda)$. The characteristic equation (\ref{equ:70}) reduces to
\begin{equation}\label{equ:71}
\big (1-T_{11}-T_{12}\big )\big (1-T_{11}+T_{12}\big )=0
\end{equation}
or equivalently to~\cite{mirollo2007the}
\begin{equation}\label{equ:72}
1=\hat{g}\frac{a(\omega)+b(\omega)}{\lambda-c(\omega)},
\end{equation}
\begin{equation}\label{equ:73}
1=\hat{g}\frac{a(\omega)-b(\omega)}{\lambda-c(\omega)}.
\end{equation}

These equations require the consideration of two different scenarios. Firstly, for $q<1$, the equations reduce to
\begin{equation}\label{equ:74}
\bigg (\frac{1}{K}\bigg )^{\frac{1}{\beta}}=\hat{g}\frac{\lambda \beta |\omega|q^{-\frac{\beta+1}{\beta}}\big (1-\sqrt{1-q^2}\big )}{\lambda^2+\omega^2(1-q^2 )},
\end{equation}
and
\begin{equation}\label{equ:75}
\bigg (\frac{1}{K}\bigg )^{\frac{1}{\beta}}=\hat{g}\frac{\lambda  |\omega|q^{-\frac{\beta+1}{\beta}}\big (q^2-1+\sqrt{1-q^2}\big )}{\lambda^2+\omega^2(1-q^2)}.
\end{equation}
According to Eq.~(\ref{equ:08}), the only solution to the self-consistency equation is $R=q=0$. For $\beta<1$, $\lim\limits_{q \to 0}q^{-\frac{\beta+1}{\beta}}(1-\sqrt{1-q^2})=\lim\limits_{q \to 0}q^{-\frac{\beta+1}{\beta}}(q^2-1+\sqrt{1-q^2})=\lim\limits_{q \to 0}\frac{1}{2}q^{\frac{\beta-1}{\beta}}=+\infty$, which implies that the incoherent state loses its stability once $K>0$. This in fact corresponds to the so-called vanishing onset. Moreover, for $\beta>1$, $\lim\limits_{q \to 0}q^{-\frac{\beta+1}{\beta}}(1-\sqrt{1-q^2})=\lim\limits_{q \to 0}q^{-\frac{\beta+1}{\beta}}(q^2-1+\sqrt{1-q^2})=\lim\limits_{q \to 0}\frac{1}{2}q^{\frac{\beta-1}{\beta}}=0$. In order to balance Eqs.~(\ref{equ:74}) and (\ref{equ:75}), $\lambda=\pm i\omega$ must correspond to a continuous spectrum, implying that the incoherent state remains neutrally, or asymptotically, stable for all $K>0$. Particularly, for $\beta=1$ $\lim\limits_{q \to 0}q^{-\frac{\beta+1}{\beta}}(1-\sqrt{1-q^2})=\lim\limits_{q \to 0}q^{-\frac{\beta+1}{\beta}}(q^2-1+\sqrt{1-q^2})=\lim\limits_{q \to 0}\frac{1}{2}q^{\frac{\beta-1}{\beta}}=\frac{1}{2}$, so that thus Eqs.~(\ref{equ:74}) and (\ref{equ:75}) degenerate to the same characteristic equation~\cite{strogatz1991stability}
\begin{equation}\label{equ:76}
\frac{2}{K}=\hat{g}\frac{\lambda|\omega|}{\lambda^2+\omega^2},
\end{equation}
which is consistent with the results obtained in Ref.~\cite{xu2018origin}. The incoherent state loses its stability  at a critical point $K_f$ which is determined by the condition $\rm{Re}(\lambda) \to 0^{+}$.

Secondly, and lastly, for $q>1$, Eqs.~(\ref{equ:74}) and (\ref{equ:75}) correspond precisely to the limit formula of Eqs.~(\ref{equ:46}) and (\ref{equ:47}), respectively, where $\hat{g}\to \frac{1}{N}\sum_{i=1}^N$, which thus completes the proof.

\begin{figure*}
\centering
\footnotesize
\includegraphics[width=\linewidth]{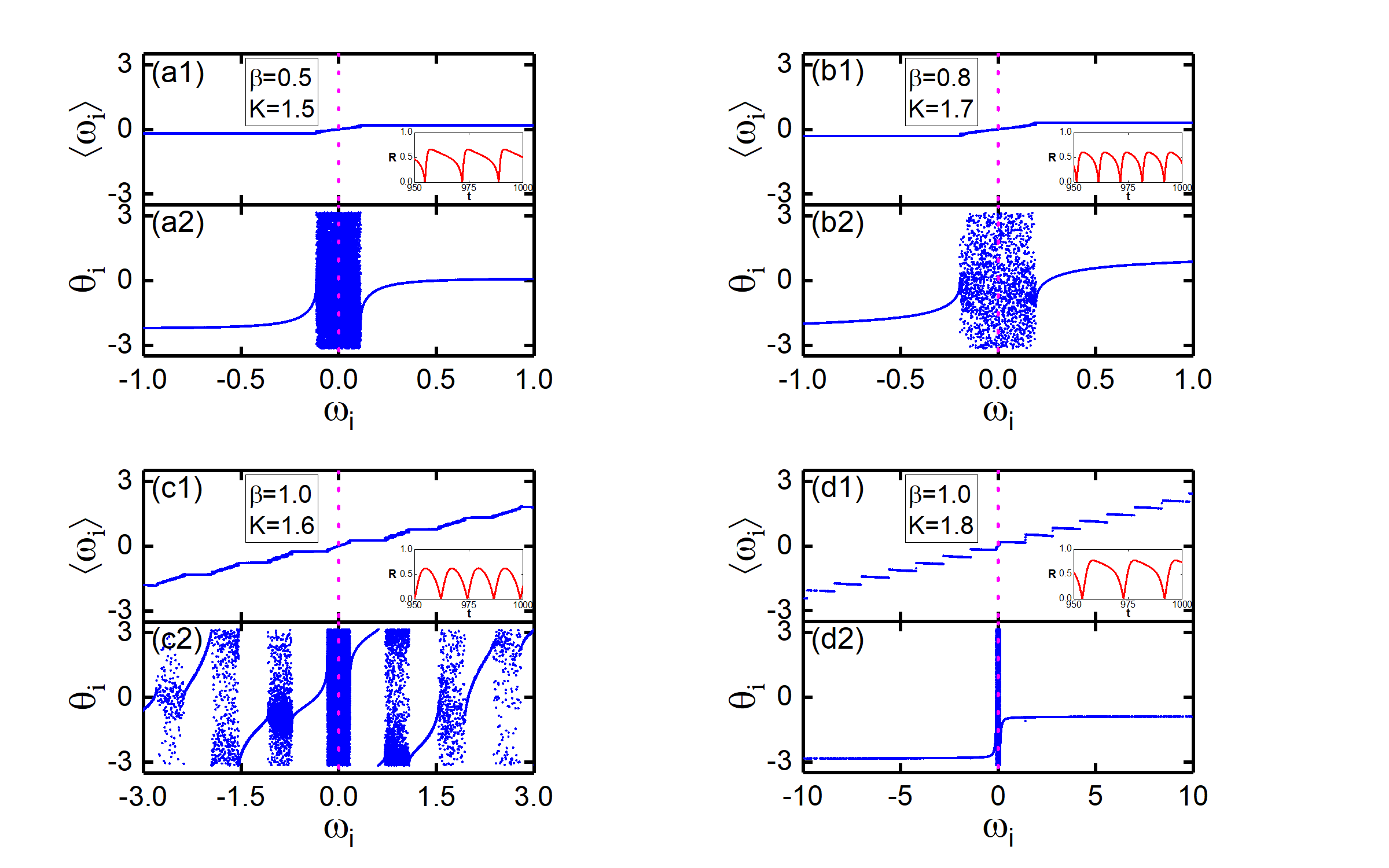}\\
\caption{Microscopic properties of the different oscillatory states. The top columns show the instantaneous phases in dependence on the natural frequencies, while the bottom columns show the effective frequencies $\left\langle \omega_i\right\rangle=1/T\int_{t}^{t+T}\dot{\theta}_i(t)dt$ in dependence on the natural frequencies. The inserts show the time series of the order parameter. The following parameters have been used: in (a1) and (a2) $\beta=0.5$, $g(\omega)=1/2$ with $\omega \in (-1,1)$, $K=1.5$, in (b1) and (b2) $\beta=0.8$, $g(\omega)=1/2$ with $\omega \in (-1,1)$, $K=1.7$, in (c1) and (c2) $\beta=1.0$, $g(\omega)=\frac{\Delta}{2\pi}\Big [\frac{1}{(\omega-\omega_0)^2+\Delta^2}+\frac{1}{(\omega+\omega_0)^2+\Delta^2}\Big ]$, (i.e., $g(\omega)$ is a bimodal Lorentzian distribution with $\pm \omega_0$ being the locations of the two peaks of the distribution, and $\Delta$ is the half width of the distribution ) with $\omega_0=0.3, \Delta=0.1$, $K=1.6$, and in (d1) and (d2) $\beta=1.0$, the bimodal Lorentzian distribution $g(\omega)=\frac{\Delta}{2\pi}\Big [\frac{1}{(\omega-\omega_0)^2+\Delta^2}+\frac{1}{(\omega+\omega_0)^2+\Delta^2}\Big ]$, with $\omega_0=0.3, \Delta=0.1$, $K=1.8$. For the simulations, we have used the same values for $N$, $t$, $\Delta t$, and $T$ as in Fig.~\ref{fig:01}.}
\label{fig:02}
\end{figure*}

\section{Numerical results}
\label{sec:06}

To show how the heterogeneous and nonlinear coupling affects the phase coherence, we study the system numerically. Figs.~\ref{fig:01} and \ref{fig:02} illustrate the macroscopical and microscopical synchronization properties, respectively. When the exponent $\beta$ is fixed, and the natural frequencies are drawn independently from a symmetric distribution $g(\omega)$, we can indeed observe rich collective dynamics from the still relatively simple extension of the traditional Kuramoto model. In Fig.~\ref{fig:01}, we show the time-averaged amplitude of the order parameter, $\left\langle R\right\rangle=\frac{1}{T}\int_{t}^{t+T}R(t)d t$ with $T$ being the time-averaged window, as a function of the global coupling strength $K$. In the simulations, both directions for synchronization transitions are considered, i.e., $K$ is first increased adiabatically from $0$ and then decreased. The presented simulation results reveal that the inhomogeneity and nonlinearity in the coupling combines together to affect the overall collective dynamics of the system, leading to a number of interesting dynamical phenomena.

In particular, for $\beta=1$ (left column), the system displays three types of phase transitions to synchronization depending on the difference between $K_f$ and $K_c$. Specifically, if $K_f>K_c$ [Fig.~\ref{fig:01}(a1)], the system exhibits a first-order (explosive) phase transition to synchronization. As the system transitions from the completely incoherent state ($\left\langle R\right\rangle\sim 0$) to the fully locked state ($\left\langle R\right\rangle\sim 1$) at the critical coupling $K_f$ under forward continuation, when $K$ is gradually decreased, another abrupt transition from synchronization to incoherence occurs at $K_c$. In the hysteretic region, $K\in (K_c,K_f)$, the system admits a bistability, where the stable incoherent state and synchronized state coexist. If $K_f=K_c$ [Fig. ~\ref{fig:01}(a2)], the system undergoes a hybrid phase transition, which is characterized by a vanishing hysteresis loop. Interestingly, if $K_f<K_c$ [Fig.~\ref{fig:01}(a3)], a tiered phase transition to synchronization occurs, where the system converts from the incoherent state to the fully synchronized state, which is mediated by an oscillatory state that emerges when $K\in(K_f, K_c)$.

When the exponent $\beta\neq 1$, the phase diagrams change considerably. For $\beta<1$ (middle column), we have shown that $K_f=0<K_c$, so that the system displays a tiered phase transition to synchronization with a vanishing onset, where the incoherent state is replaced by an oscillatory state the amplitude of the mean-field is time-varying that arises in the region $K\in (0, K_c)$. However, for $\beta>1$ (right column), $K_f=+\infty>K_c$, an irreversible abrupt desynchronization transition is observed, featuring a hysteresis area, whose width is infinite. The system experiences a discontinuous transition from the fully locked state to the incoherent state at $K_c$ in the backward continuation, where $\left\langle R\right\rangle $ jumps from $R_c$ to $0$ directly. In contrast, there is no corresponding counterpart of this abrupt transition when going from the incoherent to the coherent state.

In Fig.~\ref{fig:02}, we highlight the oscillatory states that emerge for the intermediate region of coupling strengths~\cite{note1}. We show the instantaneous phases, effective frequencies, and the time series of $R(t)$. These results illustrate that, in the oscillatory state, the oscillators form coexisting synchronized and drifting groups, or spontaneously partition into different clusters according to their natural frequencies. Moreover, the effective frequencies remain entrained in each cluster, but they do not synchronize with one another. As a result, the instantaneous phases show a correlation at a fixed time, thereby causing the order parameter to behave periodically. For further details, we refer to the results shown in Fig.~\ref{fig:02} and the corresponding description of the parameter values in the figure caption.

\section{Discussion}
\label{sec:07}

In summary, we have studied an extension of the Kuramoto model, where the randomness of the coupling strength is related to the natural frequencies of the coupled oscillators, and the mean-field is modified to depend on the global coherence of the system. In doing so, we have thus jointly considered heterogeneity and nonlinearity in the coupling of the Kuramoto model, which has led us to uncover a number of fascinating dynamical states and transitions associated with their emergence. These include coherent and incoherent states, two-cluster states, and oscillatory states, which can emerge via various transitions among them, including explosive synchronization transitions, hybrid transitions with hysteresis absence, abrupt irreversible desynchronization transitions, and tiered phase transitions with or without a vanishing onset.

Despite the wealth of different dynamical states and the transitions between them, our model still proved to be analytically tractable, thus allowing us to provide firm theoretical foundations for the observed collective behavior. In particular, we have provided a rigorous stability analysis of the equilibrium states by studying their eigenspecturm properties in the finite and infinite size limit. Moreover, we have shown that the critical points that correspond to bifurcations of steady states and the stable conditions for their occurrence in the phase space can be obtained by means of the matrix theory and the Ott-Antonsen reduction.

Although the Kuramoto model and its many variants have been studied extensively in the past decades, we have shown that the joint consideration of heterogeneity and nonlinearity in the coupling among oscillators affords new insights into the emergent collective dynamics that could be of relevance for complex systems where such conditions apply. Previous research has already highlighted the importance of heterogeneous interactions for the dynamics of excitatory-inhibitory neurons, and for the emergence of consensus amongst conformists and contrarians in different social settings~\cite{cb2003, ms2005, 2008}. Heterogeneity and nonlinearity are also often related to plasticity, which leads to changes in the connection strength among individuals to achieve efficient global states. This may have implications in neuroscience as well as in the social sciences, where such interaction scenarios often unfold. More generally, our research reveals how an alternative coupling scheme can lead to the emergence of diverse rhythmic states in complex systems that exhibit synchronization transitions with relatively small changes to system parameters. Since such complexity is rarely analytically tractable, we hope that our model will prove to be inspirational for future research along these lines and find applicability also in realistic complex systems.

\begin{acknowledgments}
This work is supported by the National Natural Science Foundation of China (Grants No. 11905068, 11875132, and 11835003), the Scientific Research Funds of Huaqiao University (Grant No. ZQN-810), the Natural Science Foundation of Shanghai (Grants No. 18ZR1411800), and the Slovenian Research Agency (Grant Nos. P1-0403 and J1-2457).
\end{acknowledgments}


\begin{thebibliography}{99}
\bibitem{note3} We emphasize that the form of the coupling that we are considering strongly deviates from the classical Kuramoto model, the huge body of literature quoted above points actually to the fact that a such form of coupling (or similar ones which include a dependence on the oscillators’ natural frequencies) are in fact of physical interest, and may be considered as a reasonable hypothesis for the description of collective dynamics observed in real-world systems.
\bibitem{note4} We remark that conformity is a well-established social psychology phenomenon demonstrated through experiments, which showed that people could conform to a majority to gain the social approval of others. Individual behavioral responses that vary with the others’ behavior are also commonly seen in animals. For example, social insects such as Temnothorax ants and honeybees select their new nest site based on a popularity competition of recruitment between potential nests. These examples suggest that individual behavioral response to conform or synchronize with others depends on the degree of consensus or synchronization of the group at that moment. From the point of view of coupled phase oscillator models, one may use NC to better model conformity behavior.
\bibitem{note} In principle, a nonzero $\Omega$ corresponding to the traveling waves should be considered for a complete self-consistent analysis. However, in this paper we do not discuss this scenario. On the one hand, the symmetry of the system ensures that $\Omega=0$ holds for the self-consistent equations. On the other hand, numerical simulations do not find the evidence of the traveling waves.
\bibitem{note1} Throughout this paper, the oscillatory states represent the collective states, in which the amplitude of the order parameter is time-dependent.
\end{thebibliography}
\end{document}